\documentclass[twocolumn,showpacs,preprintnumbers,amsmath,amssymb]{revtex4}

\usepackage{graphicx}
\usepackage{dcolumn}
\usepackage{bm}

\begin{document}


\title{Geometrical phase of thermal state in hydrogen atom}
\author{Guo-Qiang Zhu}
 \affiliation{Zhejiang Institute of Modern Physics, Zhejiang University, Hangzhou, P.R. China}


\begin{abstract}
In this paper, the geometric phase of thermal state in hydrogen
atom under the effects of external magnetic field is considered.
Especially the effects of the temperature upon the geometric phase
is discussed. Also we discuss the time evolution of entanglement
of the system. They show some similar behaviors.
\end{abstract}

\maketitle

\section{Introduction}
The concept of geometric phase was first introduced by Panchartnam
in his study of interference of classical light in distinct states
of polarization \cite{pancharatnam}. Berry's work showed a quantum
pure state can retain the information of its motion when it
undergoes a cyclic evolution \cite{berry}. Simon \cite{simon}
subsequently recasted the mathematical formation of Berry phase with
the language of differential geometry and fibre bundles. He observed
that the origin of Berry phase is attributed to the holonomy in the
parameter space. It has been pointed out that the non-Abelian
holonomy may be used in the construction of universal sets of
quantum gates for the purpose of achieving fault-tolerant quantum
computation \cite{zanardi1,zanardi2}. The holonomy quantum
information processing is vary important due to its robustness to
imperfections, such as decoherence and the random unitary
perturbations.

For a composite system, the Berry phase will be changed by the
intersubsystem couplings. Berry phase can be used to the
implementation of quantum computation; all the system for the above
purpose are composite. Composite systems have great importance in
quantum computation, such as the manipulation of qubits, the
construction of entanglement and the realization of logic
operations. X.X. Yi et al \cite{xxyi} studied the Berry phase in a
composite system with one driven subsystem and found that the Berry
phase for a mixed state $\rho(t)=\sum_j p_j|E_j\rangle\langle E_j|$
is the average of the individual Berry phases, weighted by their
eigenvalues $p_j$.

%
%

In the early discussions, most researches were focused on the
evolution of pure states. Uhlmann was probably the first to address
the issue of mixed state holonomy, but as a purely mathematical
problem \cite{uhlmann1,uhlmann2}. Later Sj\"{o}qvist et al discussed
the geometric phase for non-degenerate mixed state under unitary
evolution in his famous paper \cite{sjoqvist}, basing on the
Mach-Zender interferometer. The holonomy  attacked interested due to
its potential importance to fault-tolerent quantum information
processing. M. Nordling et al introduced the concept of non-Abelian
holonomy for adiabatic transport of energetically degenerate mixed
quantum states \cite{nordling}.

Recently Tong, \emph{et al}\cite{tong} proposed a definition for
mixed-state geometric phase which was gauge invariant and applicable
to the non-unitary evolution,
\begin{equation}
\gamma_g=\arg(\sum_{k=1}^4\sqrt{\lambda_k(0)\lambda_k(t)}\langle
k|U(t)|k\rangle e^{-\int_0^{t}\langle k(t')|\dot{k}(t')\rangle
dt'}).
\end{equation}where $\lambda_k(t)$ is the k-th eigenvalue of the
density matrix $\rho(t)$. When the Hamiltonian is independent of
time, the phase can be reduced into
\begin{equation}
\gamma_g=\arg(\sum_{k=1}^{4}\lambda_k\langle k|U(t)|k\rangle
e^{i\langle k|H|k\rangle t})\label{geometric}
\end{equation}
in which $U=e^{-iHt}$ and $|k\rangle$ is the eigenstate of the
initial density operator.

The connection between geometric phase and entanglement has now
drawn much attention. In a recent work, S. Ryu and Y. Hatsugai
\cite{ryu} tried to establish a connection between the lower bound
of the von-Neumann entropy and the Berry phase defined for quantum
ground states. It has been proved that in the XY spin chains,
geometric phase shows scaling behavior in the vicinity of quantum
phase transition point \cite{slzhu}, just like entanglement
\cite{osterloh}. Up to now, the relationship between geometric
phase and entanglement is not clear yet.
\section{Model}
In this paper, the thermal state of the hydrogen atom is discussed.
As one knows, in the hydrogen atom, the electron spin is coupled to
the nuclear spin by the hyperfine interaction. The hyperfine line
for the hydrogen atom has a measured magnitude of 1420 MHz in
frequency. Some calculation on the basis of first-order perturbation
for the magnetic dipole interaction between the electron and nucleus
gives contribution to the coupling strength of $\mathbf{I}\cdot
\mathbf{S}$ term. Taking account of the effect of an external
magnetic field, one can have a model Hamiltonian which is \cite{bec}
\begin{equation}
H=H_0+H_I
\end{equation}
where \begin{eqnarray}H_0&=&J(I_x\cdot S_x+I_y\cdot S_y+I_z\cdot
S_z),\\H_I&=&C S_z+D I_z,
\end{eqnarray}$J$ is the coupling constant. Here we have
assumed that the electronic orbital angular momentum $L$ is zero.
The parameters $C$ and $D$ are related with external magnetic
fields,
$$C=g\mu_B B, \ \ D=-\frac{\mu}{I}B.$$ It is obvious that the
commutation $[H_0,H_I]\neq0$.

As we know, for a hydrogen atom, the nucleus and electron has both
spin-1/2, $\mathbf{I}=\mathbf{S}=\frac{1}{2}\mathbf{\sigma}$. The
nuclear magnetic moment $\mu$ equals $2.793\mu_N$ where
$\mu_N=e\hbar/(2m_p)$. In general C is much larger than D,
$|C/D|\sim m_p/m_e\approx 2000$, so for most applications D may be
neglected. At the same level of approximation the $g$ factor of the
electron may be put equal to 2.

In the next sections one can discuss the geometric phase and
entanglement for hydrogen atom ${}^1 H$.
\section{Geometric Phase}
In this section, the geometric phase of mixed states for hydrogen
atom will be discussed. One can assume that at the initial time
$t=0$, the system is subject to the external magnetic field
$\mathbf{B}$, the model is not degenerate. As we know, if the
magnetic field is not imposed, the ground state is degenerate. The
external fields destroy the degeneracy completely. The initial
state is set to be a thermal state,
\begin{equation}
\rho_0=\sum_{i=1}^{4}p_i |\psi_i\rangle\langle\psi_i|.
\end{equation}
Here $p_i=\exp(-\beta E_i)/Z$ and $Z=Tr\exp\{-\beta H\}$ is the
partition function. $\beta=1/(kT)$. In fact,
\begin{equation}
\rho_0=e^{-\beta H}/Tr e^{-\beta H}.
\end{equation}
where $H=H_0+H_I$.

At that time, the four eigenvalues of the initial Hamiltonian $H$
are $(J\pm2 C)/4$, $(-J\pm\sqrt{C^2+J^2})/4$. It is obvious that
the system is non-degenerate.

\begin{figure}
\begin{center}
\includegraphics[width=6cm]{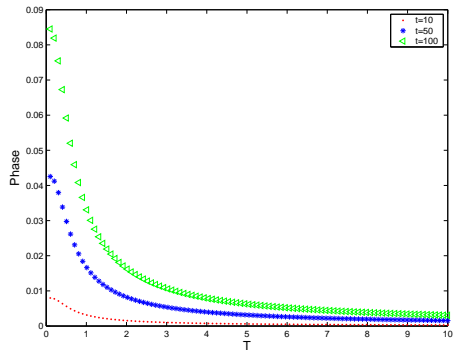}
\caption{\label{temperature} Geometric phase versus temperature at
certain time.}
\end{center}
\end{figure}

At the time t=0+, one can assume that there is a little change of
external magnetic fields, the interaction term is now
$(1+\epsilon)H_I$ and $\epsilon$ is a small constant. The
eigenstate of the model will evolve with the time. Now the time
evolution matrix becomes $U=e^{-i H' t}$, and
$H'=H_0+(1+\epsilon)H_I$ which implies that magnetic field jumps
to another fixed value at that moment. Then one calculate the
geometric phase of thermal states which involve with the time by
means of Eq.\ref{temperature}. Obviously, the geometric phase is
dependent on the external magnetic field, intersubsystem
couplings, temperature and the time.

The explicit form of geometric phase is very tedious, one only can
study some special cases. Then one can discuss the effects of some
parameters upon the geometric phase respectively.

(1). The coupling constant $C$, in fact, the external magnetic
field.

When the C approaches zero, i.e., the magnetic field vanishes, the
Hamiltonian $H$ and the density matrix $\rho$ is commutative, the
density matrix is not changed with the time. Therefore in this case
the geometric phase is zero.

\begin{figure}
\begin{center}
\includegraphics[width=6cm]{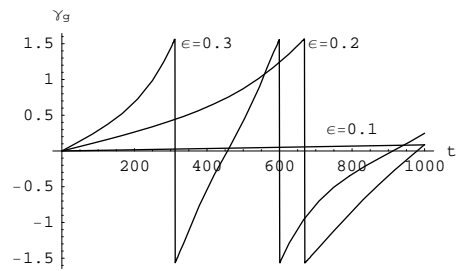}
\caption{\label{epsilon} Hydrogen, for different $\epsilon$.}
\end{center}
\end{figure}

For simplicity, we assume at the initial time, J=1, C=1, $\beta=1$.
Then one study the time evolution of geometric phase for different
coupling C, i.e., for different external magnetic field. Here one
only need to study the effect of $\epsilon$ upon the geometric
phase. The Fig.\ref{epsilon} shows that for the same initial state,
the magnetic field changes more, the geometric phase changes fast
accordingly.

\begin{figure}
\begin{center}
\includegraphics[width=6cm]{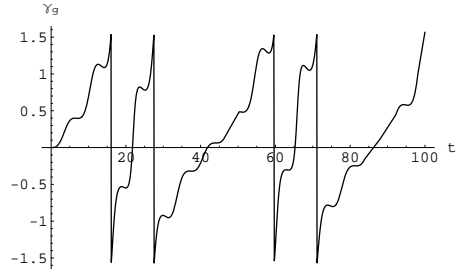}
\caption{\label{c0}For hydrogen, the external magnetic field is
disappeared. $J=1$, $\beta=1$, $C=1$.}
\end{center}
\end{figure}

One can consider a special case in which $\epsilon=-1$, which
means at that moment the external magnetic field is disappeared.
The geometric phase still evolve with the time. The relationship
can be plotted in the Fig.\ref{c0}.

(2). The coupling constant $J$.

\begin{figure}[ht]
\begin{center}
\includegraphics[width=6cm]{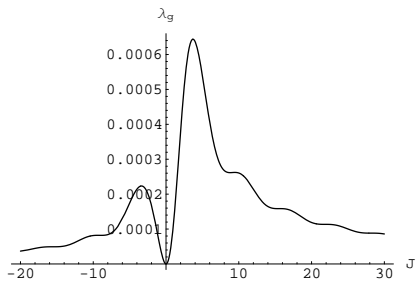}
\caption{\label{J} For hydrogen atom, geometric phase versus J, when
$C=1$, $\beta=1$ at the time $t=1$.}
\end{center}
\end{figure}

The Fig.\ref{J} shows at certain time $t=1$, if $J=0$, the geometric
phase $\gamma_g$ vanishes. For $J>0$ and $J<0$, the behaviors are
not same. When $J>0$, $\gamma_g$ can reach a larger maximum in the
process of evolution than the case of $J<0$. Obviously, when $|J|$
approaches infinity, geometric phase is vanishing. It is obvious
that in this case,the term $H_0$ will dominate the whole
Hamiltonian, the density matrix will approaches zero.

(3). The temperature parameter $T$.

\begin{figure}
\begin{center}
\includegraphics[width=6cm]{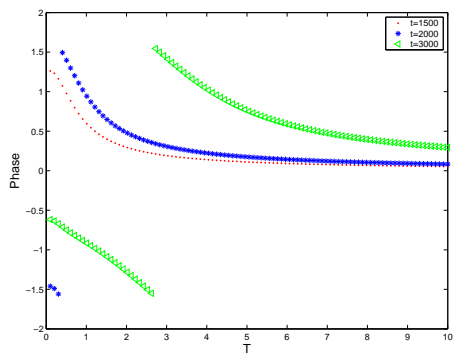}
\caption{\label{temp2} Geometric phase versus temperature.}
\end{center}
\end{figure}

One also can study the the relationship between geometric phase and
the temperature at  certain time. The results can be plotted in the
Fig.\ref{temperature}.
From the figure, one can see at certain time, the geometric phase
vanishes when the temperature is relatively high.  One can see the
details in the Fig.\ref{temp2}.
When the temperature is increased, the system becomes more mixed
and disordered. It drivers the geometric phase to approach zero.
\section{Concurrence}
In the above, we have discussed the time evolution of geometric
phase. As we know, entanglement may be created via the interaction
of jointed measure, therefore the way in which intersubsystem
couplings changes, the geometric phase of a composite system and
those of the subsystem is of interest.  Here, we shall study the
time evolution of entanglement. As we know, for a bi-qubit system,
one can use concurrence to measure the entanglement. Speaking
briefly, for a bipartite system with the density matrix $\rho$, one
can define concurrence
\begin{equation}
\mathfrak{C}=\max\{0,\lambda_1-\lambda_2-\lambda_3-\lambda_4\},
\end{equation}
where $\lambda_i$ is the eigenvalue of the matrix
$\sqrt{\rho\sigma_y\otimes\sigma_y\rho^{\ast}\sigma_y\otimes\sigma_y}$
and $\lambda_1$ is the biggest one. The system is unentangled
(maximally entangled) when the concurrence $\mathfrak{C}$ is 0 (1)
\cite{wootters1,wootters2}. Later, the concurrence was used to
study the entanglement of the spin chains at finite temperature
when the density is the Gibbs density matrix $\rho=e^{-H/kT}/Z$
and $Z=tr e^{-H/kT}$ \cite{arnesen}.

\begin{figure}
\begin{center}
\includegraphics[width=6cm]{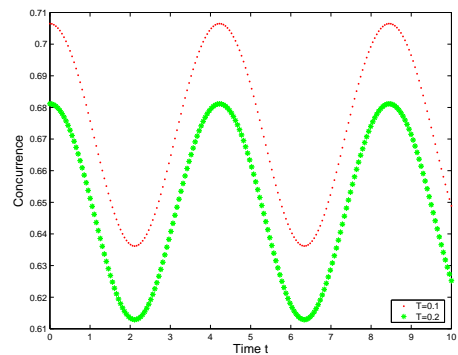}
\caption{\label{con-time} Concurrence versus time at different
time.}
\end{center}
\end{figure}

\begin{figure}
\begin{center}
\includegraphics[width=6cm]{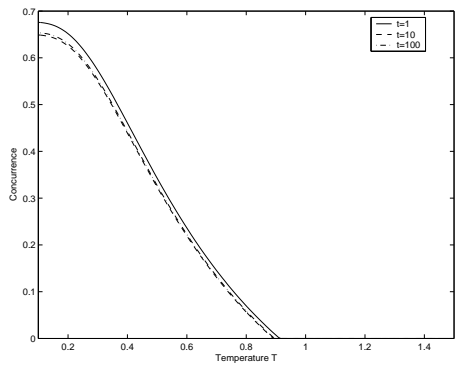}
\caption{\label{con-t} The time evolution of Concurrence versus
temperature.}
\end{center}
\end{figure}

In our work, the density matrix is dependent on the time,
\begin{equation}
\rho(t)=e^{-iHt}\rho_0 e^{iHt}.
\end{equation}
The result is a bit complicated. Fig.\ref{con-time} shows that
with the time increasing, the concurrence changes like wave with
the time. The entanglement do not vanishes as the time approaches
infinity.
 Fig.\ref{con-t} shows that as the temperature becomes lower, the value of concurrence
becomes larger, which implies the system is more entangled at lower
temperature. When the temperature is higher than a threshold value,
the concurrence vanishes. It implicates that in this case the system
is not entangled.

If one compares the properties of the temperature evolution of
geometric phase and concurrence respectively, one can see there are
some similar properties of them. At certain time, the two both
approach zero as the temperature increases. At relatively low
temperature, they both decrease with the temperature, which implies
there are some relationship between the geometric phase and
entanglement. For a relatively large temperature interval, geometric
phase is varying very slowly with temperature.

\section{Summary}
We has discussed the time evolution of geometric phase of the
hydrogen atom. The relationship between geometric phase and some
parameters, such as temperature, external field and coupling
constant, has been obtained. We also discussed the time evolution
of entanglement. There are some similar properties of their
behaviors.

The helpful discussions with Zhe Sun is acknowledged. The work is
supported by NSF-China under grant No. 10405019, 10225419 and
90103022.

\end{document}